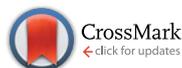

# A liquid contact line receding on a soft gel surface: dip-coating geometry investigation

Tadashi Kajiya,[ab] Philippe Brunet,[a] Laurent Royon,[a] Adrian Daerr,[a] Mathieu Receveur[a] and Laurent Limat[a]

We studied the dynamics of a liquid contact line receding on a hydrophobic soft gel (SBS–paraffin). In order to realize a well-defined geometry with an accurate control of velocity, a dip-coating setup was implemented. Provided that the elastic modulus is small enough, a significant deformation takes place near the contact line, which in turn drastically influences the wetting behaviour. Depending on the translation velocity of the substrate, the contact line exhibits different regimes of motions. Continuous motions are observed at high and low velocities, meanwhile two types of stick–slip motion – periodic and erratic – appear at intermediate velocities. We suggest that the observed transitions could be explained in terms of the competition between different frequencies, i.e., the frequency of the strain field variation induced by the contact line motion and the crossover frequency of the gel related to the material relaxation. Our results provide systematic views on how the wetting of liquid is modified by the rheological properties of a complex soft substrate.

## I. Introduction

Gels are intriguing states of matter which consist of a three-dimensional cross-linked network expanded throughout their whole volume by a fluid.[1–3] Due to their unique mechanical and osmotic properties, gels have numerous applications ranging from medical science[4,5] (e.g. soft contact lens and artificial cartilage) to pharmaceutical and food chemistry (e.g. drug delivery and food capsuling).[6–8] For these applications, the key is the proper characterization and tuning of the gel interfaces, as they determine the friction,[8,9] adhesion,[10] surface tension[11] and wetting.[12] Also, from a viewpoint of fundamental questions on wetting, gels are interesting model systems to study how singular the liquid wetting behaviour is, compared to that on infinitely hard solids.[13,14]

Wetting on a soft surface (of elastic modulus $G$ smaller than 100 kPa) has been investigated for decades,[15–18] and is still attracting great attention. In order to provide a clear view on how the three phase contact line behaves on highly deformable surfaces, active investigations are carried out on various types of materials such as gels, elastomers, and thin polymer films.[19–22] Experiments on spreading,[23] evaporation,[24,25] condensation and icing[26,27] of tiny sessile droplets on those surfaces have been realized. According to the primitive modelling by de Gennes and Shanahan,[28] the "elasto-capillary deformation" plays an important role. The liquid surface tension $\gamma$ pulls up the substrate, forming a surface ridge of typical spatial extent $\gamma/G$.

In the case of wetting on gel surfaces, their complex rheological properties largely affect the statics and dynamics of the contact line.[29] Previously, we have studied the forced spreading of water sessile drops inflated at a constant rate on hydrophobic poly(styrene-butadiene-styrene) (SBS)–paraffin gels.[30] Unlike the case of elastomers,[31] we have observed that the contact line exhibits different regimes of motions depending on the spreading velocity: two types of continuous motions and a stick–slip motion. Therein, one of the transitions seems to depend on the crossover frequency $f_{cross}$ at which the gel mechanical response changes from elastic to viscous. Similar stick–slip behaviour of the contact line was reported by Pu et al. in wetting on thin polymer films.[22,32]

Toward a general interpretation of the wetting dynamics, and especially of which length and time scales are associated with such transitions, it is necessary to develop an experiment in a well-defined geometry and with an accurate control of the contact line velocity over a wide range. Spreading of the droplet is not an ideal situation for this, as the size of the system continuously changes and the mean contact line velocity is not easily set to a constant value. Furthermore, the strongly curved geometry of the droplet surface may influence the contact line behaviour in a complex way.

In the present article, we implemented a dip-coating experiment, in which the gel substrate is withdrawn from a water bath at a constant translation velocity. Similar to the experiments on spreading droplets, the contact line exhibits two continuous motions at high and low velocities and stick–slip

[a] Laboratoire Matière et Systèmes Complexes (MSC), UMR 7057 CNRS, Université Paris Diderot, Bâtiment Condorcet, 10 rue Alice Domon et Léonie Duquet, 75205 Paris Cedex 13, France

[b] Max Planck Institute for Polymer Research, Ackermannweg 10, 55128 Mainz, Germany

motion at an intermediate velocity. In addition, we found that the stick–slip motion is distinguished into two types; large and regular stick–slip motions occurring coherently in the entire part of the contact line and small and irregular ones triggered by localized pinning spots. We discuss how the transitions are determined by the parameters such as the frequency of the strain field near the moving contact line. Finally, we propose a qualitative modelling which predicts the existing range of the stick–slip regimes, in reasonable agreement with the experiments.

## II. Experimental section

### A. Material

Poly(styrene-butadiene-styrene) (SBS)–paraffin gels were used for the gel substrates and distilled water (Milli-Q Integral; Millipore, USA) was used for the wetting liquid. SBS powders (G1682; Kraton Polymers, USA) were dissolved in paraffin (Norpar15; ExxonMobil, USA) being heated in a water bath at 90 °C. After SBS powders were completely dissolved, the solution was poured into a gel mold and was cooled down to ambient temperature. The gel mold consists of two glass plates separated by a rubber spacer. The dimension of the gel plate was 60 mm in length, 20 mm in width and $2.5 \pm 0.5$ mm in thickness. In our previous publication,[30] we have checked with an optical profiler that the typical roughness of the gel surface prepared by this procedure was of the order of 10 nm, which is far smaller than the typical length of the elasto-capillary deformation $\gamma/G'$. The concentration of the SBS polymer by mass $c_{pol}$ was varied from 8% to 25%.

The rheology of SBS–paraffin gels was measured using a strain-controlled rheometer (Physica MCR 500; Anton Paar, Austria) with a stainless steel cone plate and a flat sample cell. The diameter of the plate is 50 mm, the cone angle is 1°, and the gap between the cone and the sample cell was set to 0.052 mm. The shear strain amplitude was set to 1%. The measurements were conducted at temperature $25 \pm 0.2$ °C and frequencies from $10^{-4}$ Hz to 100 Hz.

In Fig. 1, examples of the curves of storage modulus $G'$ and loss modulus $G''$ of SBS–paraffin gels are plotted for two polymer mass concentrations: $c_{pol} = 10\%$ and 12%. The typical values of the storage modulus $G'$ at $f = 1$ Hz and crossover frequency $f_{cross}$ at which $G''/G' = 1$ are shown in Table 1 for $c_{SBS} = 8,10,12$ and 15%, respectively. Roughly speaking, when our gel is probed at a frequency $f < f_{cross}$, it behaves as a liquid, and as a solid for $f > f_{cross}$.

### B. Dip-coating setup

The dip-coating setup illustrated in Fig. 2 allows the retraction of a straight contact line over a typical width of 2 cm at a constant and well-controlled velocity. The gel substrate, which is attached to a glass plate, was mounted on a translation stage (PI ref. PLS-85 2″ SM 1 HLS), allowing for an accurate control of the velocity in a large range from $10^{-4}$ mm s$^{-1}$ to 10 mm s$^{-1}$. Note that the gel surface is oriented upward. The substrate was dipped into a water bath contained in a parallelepiped glass container of dimension $100 \times 100 \times 50$ mm. Then, it was withdrawn from the bath at a constant translation velocity $v$.

The front view of the contact line and the side view of the liquid meniscus were monitored using two CCD cameras (Model SLA1390 and PLA1000; Basler, Germany) with magnification lenses (CCTV lens; Avenir, France and Opto-zoom 70XL; Stemmer, France). To ease a clear visualization of the meniscus, the stage was inclined at $\alpha = 12°$ from the vertical axis. This inclination is small enough to not alter the observed behaviours compared to a vertical set up. The component of gravity tangent to the plate is only modified as the cosine of alpha, $i.e.$ less than $\alpha^2/2 \approx 3\%$. All measurements were conducted at room temperature $25 \pm 1$ °C and relative humidity $27 \pm 8\%$.

We finally remark that the gel creeping under the action of gravity can be neglected during the dip-coating process. The typical value of the gel viscosity $\eta_s$ is of order 100 kPa s ($c_{pol} = 10\%$). For the gel to sag away over the length scale of the liquid meniscus $l_{sag}(t) \approx 2.5$ mm, it takes $t \approx 10^4$ s, which is much longer than the duration of the experiment (This is estimated by Reynolds thinning law:[33] $l_{sag}(t) \sim \rho g e^2 t/\eta_s$, where $\rho$ is the density: 0.9 g ml$^{-1}$, $g$ the gravity constant, $e$ the gel thickness: 3 mm and $t$ the required time).

## III. Experimental results

### A. Different contact line motions: stick–slip $vs.$ continuous

Fig. 3 shows typical sequential pictures of the front view of the contact line and side view of the meniscus. The gel substrate is withdrawn from the water bath at a translation velocity $v = 0.015$ mm s$^{-1}$ and $c_{pol} = 10\%$. While the gel substrate moves up, the contact line of the water meniscus exhibits stick–slip behaviour. First, the contact line is pinned at a certain position on the gel surface (stick phase). The meniscus goes up at a same velocity as $v$, meanwhile the contact angle $\theta$ continues decreasing. When $\theta$ reaches a critical value close to $40°$, the contact line suddenly slips down over a length of order 1 mm (slip phase) and $\theta$ goes back to a value $\approx 75°$ before the onset of the stick phase. Successively, the contact line repeats this stick–slip motion.

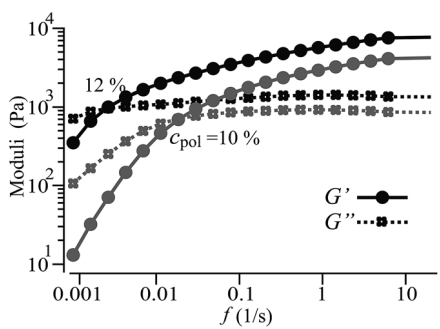

**Fig. 1** Storage $G'$ and loss $G''$ moduli of SBS–paraffin gels. Curves for two polymer mass concentrations 10% and 12% are shown.

**Table 1** Values of storage modulus $G'$ at 10 Hz and crossover frequency $f_{cross}$ where $G''/G' = 1$ for SBS–Paraffin gels of different polymer concentrations

| $c_{pol}$ (%) | 8 | 10 | 12 | 15 |
|---|---|---|---|---|
| $G'$ (kPa) | 2.99 | 4.38 | 8.00 | 15.1 |
| $f_{cross}$ (s$^{-1}$) | $5.9 \times 10^{-2}$ | $1.5 \times 10^{-2}$ | $3.0 \times 10^{-3}$ | $1.1 \times 10^{-3}$ |

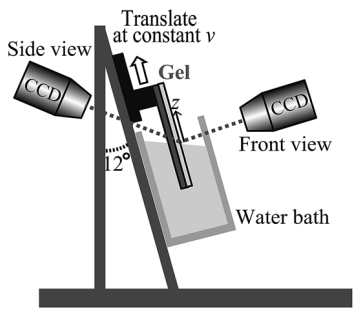

Fig. 2 A schematic of the setup for the dip-coating experiment. The substrate is fixed on a translation stage. After dipping into a liquid bath, the substrate is withdrawn at a constant velocity. The contact line was monitored with two CCD cameras; one looking from the front, the other from the side.

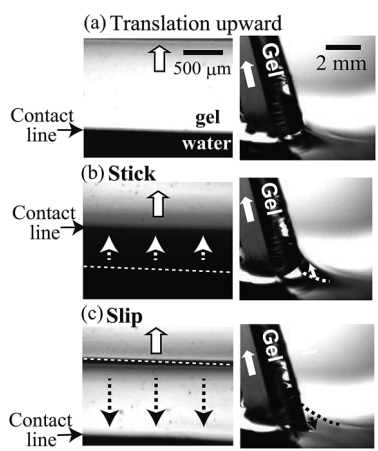

Fig. 3 Sequential pictures of the contact line on a gel of $c_{pol} = 10\%$ at translation velocity $v = 0.015$ mm s$^{-1}$. Front and side views are shown. While the gel substrate is withdrawn from the water bath, the contact line exhibits (b) stick and (c) slip motion.

In order to look at quantitatively, dip-coating experiments were conducted at different translation velocities $v$. From the sequential pictures, the position of the contact line $z$ and the contact angle $\theta$ were measured as a function of time. (Note: as shown in Fig. 2, the $z$ axis is taken parallel to the substrate, which is slightly inclined $\alpha = 12°$ from the vertical axis. The actual height of the meniscus is thus $z \cos \alpha$.)

Fig. 4 shows plots of $z$ and $\theta$ on gels of $c_{SBS} = 10\%$ at three different translation velocities: (a) $v = 3$ mm s$^{-1}$, (b) 0.01 mm s$^{-1}$, and (c) 0.001 mm s$^{-1}$. The value $z$ was obtained by tracing the position of the contact line in several vertical sections in the front view of the meniscus, then by averaging the measured values. For a large meniscus increasing to high $z$, the contact angle $\theta$ was measured directly from the picture. For a small meniscus in which one cannot directly measure $\theta$, $\theta$ was derived

from $z$ by using the formula of the meniscus profile on an inclined substrate,[34]

$$\sin(\theta + \alpha) = 1 - (z/2l_{cap})^2, \tag{1}$$

where $l_{cap} = \sqrt{\rho g / \gamma} \approx 2.5$ mm is the capillary length of water and $\alpha = 12°$ is the inclination angle of the plate.

The contact line remarkably changes its behaviour with the change of the translation velocity $v$:

• When the substrate moves up at a high velocity (e.g. $v = 3$ mm s$^{-1}$), the values of the contact line position $z$ and of the contact angle remains constant: $z \approx 2$ mm, $\theta_r \approx 75°$ (Fig. 4(a)). It indicates that during the withdrawn process, the contact line recedes continuously on the moving gel surface at a constant receding angle. When we look at the gel surface after the dip-coating process, entirely removing the substrate from the bath, the surface looks flat and no visible trace remains.

• At an intermediate velocity ($v = 0.01$ mm s$^{-1}$), the contact line exhibits the stick–slip behaviour as explained previously (Fig. 4(b)). After the dip-coating process, the gel surface shows a series of stripes of straight lines corresponding to the successive positions of the meniscus.

• At a low velocity ($v = 0.001$ mm s$^{-1}$), again the contact line recedes continuously (Fig. 4(c)). While the contact line recedes, the contact angle remains almost the same value $\theta_r$. However, $\theta_r$ is slightly lower than that of continuous regime at high $v$. After the dip-coating process, the meniscus leaves a single large ridge on the gel surface.

## B. Two types of stick–slip motions at intermediate $v$

If we carefully look at the contact line behaviour at intermediate velocities (0.003 mm s$^{-1} \leq v \leq 0.5$ mm s$^{-1}$ for gels of $c_{pol} = 10\%$), the stick–slip motion can be classified into two types: regular and irregular. Fig. 5 illustrates the major differences between these two stick–slip motions.

When the translation velocity is relatively slow ($v = 0.03$ mm s$^{-1}$ in Fig. 5(a)), the stick–slip motion is regular and periodic. The duration of one stick phase, the variation of the contact angle during the stick phase, and the length of the slip are rather similar for each repetitive event. In Fig. 5(a) for instance, the contact angle varies from $\theta_r \approx 75°$ to $\theta_{r'} \approx 40°$ during the stick phase, then successively the contact line slips over a length of 1 mm. At this translation velocity, the stick–slip motions are also spatially correlative. When we compare the position of the contact line at different points (i) and (ii) which are separated by a 1 mm distance, it is observed that each stick and slip motions nearly occurs simultaneously (with respect to the time between each jump). The slip motion on this sample rapidly propagates laterally along the contact line, and the contact line at point (ii) slightly lags from (i) to less than one second.

On the other hand, when the translation velocity is large enough ($v = 0.06$ mm s$^{-1}$ in Fig. 5(b)), the stick–slip motion becomes irregular. In this regime, the contact line exhibits small stick–slip motions, but the duration of the stick phase and the slippage length both vary largely in each jump. Also, no obvious strong spatial correlation between consecutive jumps could be noticed. Having a look at the sequential images in

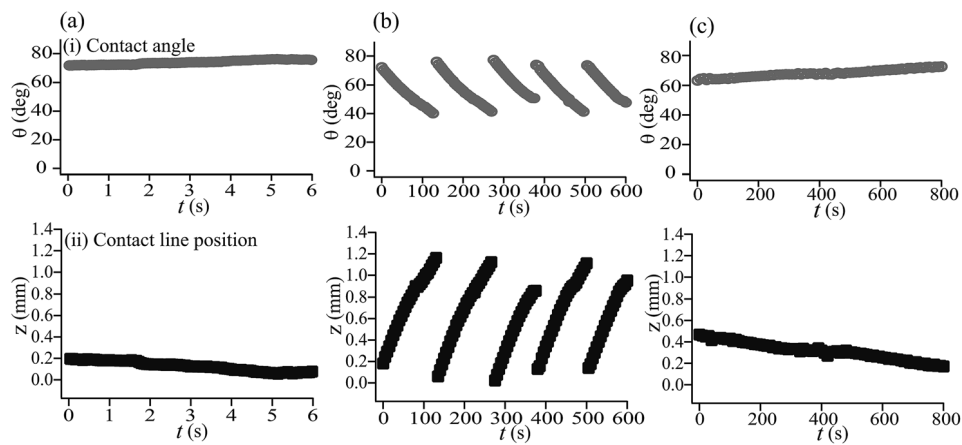

**Fig. 4** Plots of the contact angle $\theta$ and of position of the contact line $z$ against time. The gel of $c_{pol} = 10\%$ is used, and the data at three different translation velocities are shown. The velocities are (a): $v = 3$ mm s$^{-1}$, (b): 0.01 mm s$^{-1}$ and (c): 0.001 mm s$^{-1}$, respectively.

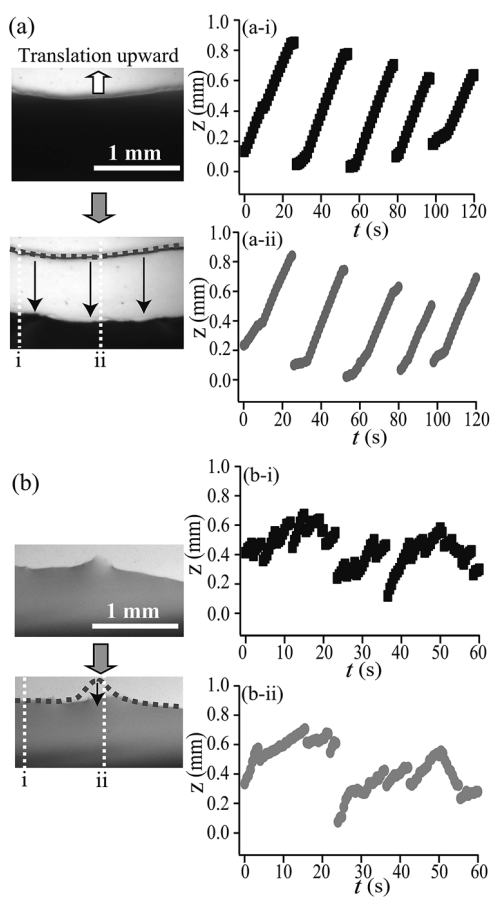

**Fig. 5** Two different types of stick–slip motions on gels of $c_{pol} = 10\%$. (a) Regular and periodic regime observed at $v = 0.03$ mm s$^{-1}$. The period of each stick–slip is almost constant. Looking at the contact line motions in different points (i) and (ii) (as shown in the inset (a-i) and (a-ii)), they also show strong spatial correlation. The different points (i) and (ii) slip at almost the same time. (b) Irregular and erratic regime observed at $v = 0.06$ mm s$^{-1}$. The period and distance of each stick–slip are irregular. Comparing the motions at different points (i) and (ii) (inset (b-i) and (b-ii)), they have quite weak spatial correlation.

Fig. 5(b), only a small area close to point (ii) slips. Therefore, the slip motion does not reach up to point (i). Nevertheless, comparing the contour of the plots, some correlation still remains, suggesting a local mechanism (possibly defect-induced). This effect is similar to the classical pinning transition that has been observed in the wetting process on rough solid surfaces.[36,37] Here, the spatial range can extend to nearly 400 μm (hence much wider than the typical size of a gel surface roughness).

### C. Classification of contact line motions against $v$ and $c_{pol}$

In order to clarify the transition between different regimes, the motions of the contact line were measured as a function of translation velocity $v$ and of polymer concentration in the gel $c_{pol}$, the latter ruling (among other mechanical properties) the elastic modulus and crossover frequency $f_{cross}$ (see Table 1). Fig. 6 shows a diagram of the contact line motions *versus* parameters $v$ and $c_{pol}$.

As the value of $v$ shifts from high to low, the contact line exhibits the four regimes of motions. (i) Continuous motion at high $v$: the entire part of the contact line recedes smoothly. The contact line pinning by local defects is not observed within the resolution of the CCD camera. (ii) Irregular stick–slip motion: a part of the contact line is locally pinned. Then after certain duration, the pinning spot is released. Precisely, we determined the onset of the irregular stick–slip motion at which the local pinning spot appeared on a scale larger than 10 pixels in the acquired image (30 μm in the real scale). (iii) Regular stick–slip motion: the pinning is not localized. The contact line is pinned over the entire width of the gel, then slips coherently. Looking at the temporal evolution, the stick–slip occurs periodically. (iv) Continuous motion at low $v$: the contact line behaviour is the same as that in regime (i).

Comparing the data for different polymer concentrations, it is shown that the critical velocities largely depend on $c_{pol}$: the softer the gel, the higher the velocity threshold for the transition from continuous to stick–slip. For instance, the threshold between the regimes (i)/(ii) is $v \approx 0.5$ mm s$^{-1}$ for gels of $c_{pol} = 10\%$ and $v \approx 1$ mm s$^{-1}$ for gels of $c_{pol} = 8\%$. However, on a rigid gel of higher polymer concentration, the critical velocities for transition shift to a considerably small value, and can sometimes even not be measured. Especially in the case of $c_{pol} = 25\%$, the continuous motion covers the entire range of $v$ in our experiments.

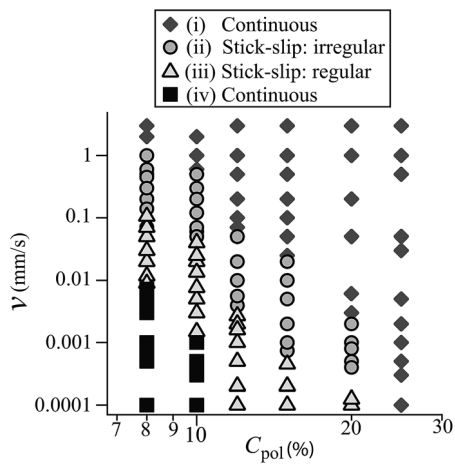

**Fig. 6** Phase diagram of the contact line motions as a function of the translation velocity $v$ and gel polymer concentration $c_{pol}$. The contact line behaviour is distinguished by 4 different regimes. (i) continuous motion at high $v$, (ii) irregular stick–slip, (iii) regular stick–slip, and (iv) another continuous motion at low $v$.

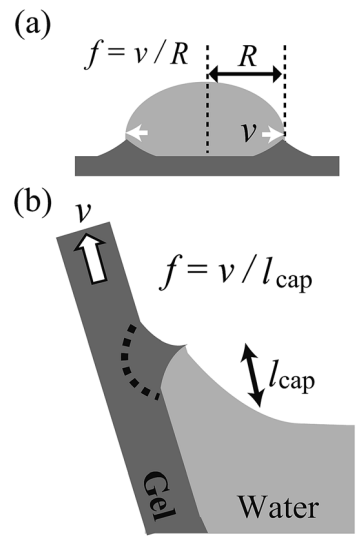

**Fig. 7** Qualitative comparison of our geometry with that of ref. 30. In both cases, the typical scale of the gel deformation is the largest scale available: (a) the radius $R$ for the droplet geometry. (b) The meniscus size $\approx l_{cap}$ for the dip-coating geometry.

## IV. Discussion

In this section, we discuss the mechanism of different regimes of contact line motions by comparing two types of frequencies, then we give a simple modelling to predict the condition for the stick–slip regime.

### A. Interpretation by the characteristic frequency of strain field

As shown in Table 1, SBS–paraffin gel has a low elastic modulus of several kPa and is thus highly deformable. The gel surface is pulled up by the unbalanced Young's force exerted in the vicinity of the contact line normal to the substrate, resulting in the formation of a sharp ridge (Fig. 7).[19–21] By analogy with the wetting of drops on elastomers,[28] the profile of the gel surface ridge is initially determined by the balance between the interfacial tensions and the elastic resistance of the gel. This was experimentally confirmed by the extraction of the deformation profile in our previous work, see Fig. 6 in ref. 30. However, when the contact line stays at the same position over a longer time, the elastic resistance gradually relaxes, storing a growing part of irreversible deformation.[29] The typical time for material relaxation is determined as $\tau_{gel} = 1/f_{cross}$, where $f_{cross}$ is the crossover frequency at which $G''/G'$ equals to 1.

In order to compare the contact line dynamics with the mechanical response and deformation relaxation of the gel, we estimate the characteristic frequency of the deformation observed on the gel surface around the moving contact line. The characteristic frequency (or inverse time) associated with the contact line motion over a ridge deformation of scale $l$ is:

$$f = v/l \qquad (2)$$

As the surface ridge near the contact line is known to be logarithmic,[17,28] $l$ could be the largest scale available in the experiment (Fig. 7). For a micrometer or millimeter sized droplet like as in our previous experiments, $l$ was naturally chosen as the radius of the droplet $l \approx R$. In the dip-coating geometry where a large amount of liquid is in contact with the gel substrate, the most natural choice should be the scale of a liquid meniscus which is of the order of the capillary length $l \approx l_{cap} = \sqrt{\rho g/\gamma} \approx 2.5$ mm.

From the data points in Fig. 6, we calculated the characteristic frequency $f = v/l_{cap}$. Considering the ratio of $f$ to the gel crossover frequency, a dimensionless number called Deborah number is calculated as,

$$\mathrm{De} = f/f_{cross} \qquad (3)$$

and is plotted in Fig. 8. Data for $c_{pol} = 8, 10, 12$ and $15\%$ are superimposed in the graph. The value for De quantifies to the extent the material relaxes in a given time of interest $1/f$.[35] The four regimes of the contact line motions (i)–(iv) are clearly separated by De. The point De = 1, at which the two frequencies have similar values, is found just in the middle of the two continuous/stick–slip transitions (i)/(ii) and (iii)/(iv). It is also close to the critical point separating the two stick–slip regimes (ii)/(iii).

We built characteristic frequencies associated with the transitions between the different regimes and corresponding to the different velocity thresholds: $f_{i/ii}, f_{ii/iii}$ and $f_{iii/iv}$. Fig. 9 shows these characteristic frequencies as a function of $c_{pol}$. It is

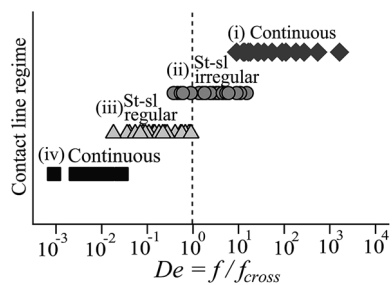

**Fig. 8** Replot of the diagram in Fig. 6 as a function of Deborah number. De is defined as the ratio of two frequencies De = $f/f_{cross}$, where $f = v/l_{cap}$ is the characteristic frequency of the strain field near the contact line and $f_{cross}$ is the crossover frequency of the gel. Data of $c_{pol}$ = 8, 10, 12 and 15% are superposed.

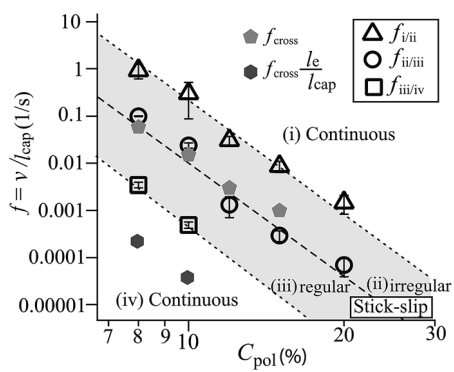

**Fig. 9** Threshold values of the characteristic frequencies $f = v/l_{cap}$ between the different regimes of contact line motions (i)–(iv). $f_{i/ii}$, $f_{ii/iii}$ and $f_{iii/iv}$ are plotted as a function of gel polymer concentration $c_{pol}$. The values $f_{cross}$ and $f_{cross} l_e/l_{cap}$ are superposed as filled dots.

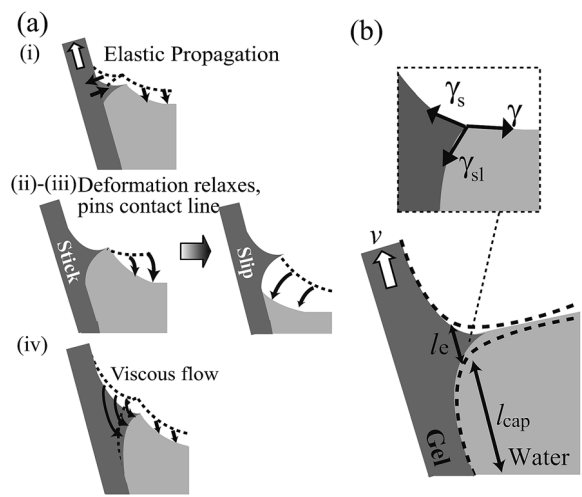

**Fig. 10** (a) Possible mechanism for the different regimes of contact line motions. The appearance of the contact line motions (i)–(iv) can be understood in terms of the competition between the frequencies $f$ and $f_{cross}$. (b) Length scales associated with the spectrum of gel surface ridge $k = 1/l$. The lower limit of the spectrum is determined by the meniscus size $l_{cap}$, while the upper limit is determined by the elasto-capillary length $l_e$ at which the logarithmic profile is truncated by the substrate surface tension $\gamma_s$.

observed that the three frequencies decrease with increasing $c_{pol}$. By comparing with the value of gel crossover frequency $f_{cross}$, all of them are proportional to $f_{cross}$. This is a strong result of the present experiments: $f_{cross}$ is the key factor that governs the three transitions observed here.

Based on the results shown in Fig. 8 and 9, the appearance of the different contact line motions can be understood in terms of the competition between the frequencies $f$ and $f_{cross}$. Fig. 10(a) shows a schematic view of a possible mechanism in each regime.

● Regime (i): when the contact line moves very fast and its characteristic frequency is far larger than the gel crossover frequency (De ≫ 1), the gel mechanical response to the contact line motion is nearly that of an elastic solid. In this regime, the ridge on the elastic surface follows exactly the motion of the contact line at which the pulling force is locally applied. The elastic deformation propagates up to a velocity of the elastic wave $\sqrt{G/\rho} \sim 1$ m s$^{-1}$, which is much higher than the velocity prescribed in these experiments. As a result, the contact line moves continuously. This is also consistent with what is observed on the surface of purely elastic materials.[31]

● Regimes (ii)–(iii): when the frequency of the contact line motion comes close to the crossover frequency (De ≈ 1), the response of the gel surface is not purely elastic. While the contact line moves over a distance proportional to the scale of surface ridge $l_{cap}$, the elastic stress starts to relax and a part of the surface deformation becomes irreversible. The surface ridge behaves somewhat as a "surface defect", which locally pins the contact line (stick phase), making it possible to reach apparent contact angles much lower than that on an infinitely stiff gel. The contact line has to jump over this defect for moving (slip phase), the slip phase corresponding to the time required for the contact-angle to recover a value close to the receding one on a flat surface.

Here, we note that the typical size of the pinned region depends on the value of De. At De > 1, the relaxation process is not fast enough to pin the entire part of the contact line. As a result, the contact line exhibits local and irregular stick–slips. At De < 1, the irreversible deformation reaches a sufficiently large scale to pin the contact line entirely. Therefore, it induces the regular and coherent stick–slip motion over the whole line.

● Regime (iv): when the contact line moves very slowly and $f$ is sufficiently low (De ≪ 1), the deformation of the gel surface totally relaxes until the contact line moves over a distance of order $l_{cap}$. In this situation, the gel substrate behaves nearly like a film of a very viscous liquid. As the contact line moves, a flow is driven below the gel surface due to the position shift of the surface force. The ridge is transported by this capillary flow, following the motion of the contact line continuously.

## B. Phenomenological modelling of the stick–slip regime window

In order to explain why the stick–slip regime appears within a finite range of $f$, we should take into account the fact that the spatial spectrum of the surface ridge $k = 1/l$ does not consist of a single mode, rather it has a continuous spectrum ranging over different length scales (Fig. 10(b)).

In coherence with the previous section, we consider that the lower limit of the spectrum associated with the ridge is of the order of the scale of liquid meniscus $k_{min} \sim 1/l_{cap}$. To have an idea of what could be the higher limit $k_{max}$, we refer to the recent theoretical studies of wetting on soft surfaces.[38–40] There, the authors proposed that the small scale truncation of the surface deformation would be ruled by the surface tension of the substrate $\gamma_s$. When $\gamma_s$ and substrate/liquid interfacial tension $\gamma_{sl}$ is taken into account, the logarithmic profile of the surface ridge is still valid,[40] but with a shift to a distance $l_e$ which is a cut-off scale for the divergence, connecting to the Neumann equilibrium generally observed in wetting of liquid on another liquid sheet,

$$\vec{\gamma} + \vec{\gamma}_s + \vec{\gamma}_{sl} = \vec{0}. \quad (4)$$

The length $l_e = \gamma_s/G$ is called the elasto-capillary length. For a SBS–paraffin gel, the typical value of $\gamma_s$ is 30 mN m$^{-1}$,[41] giving the value of $l_e \approx 6$ µm for $G = 5$ kPa for instance.

By introducing these two limits of the surface deformation spectrum, we expect that stick–slip motion is observed provided the gel crossover frequency is found between the two critical values

$$v/l_{cap} < f_{cross} < v/l_e \quad (5)$$

Eqn (5) can be rewritten in terms of the characteristic inverse time (or frequency) of the surface deformation $f = v/l_{cap}$, which yields the following condition for the stick–slip appearance,

$$f_{cross} \frac{l_e}{l_{cap}} < f < f_{cross}. \quad (6)$$

Furthermore, we remark that the crossover frequency can be re-expressed as a ratio between the elastic modulus and viscosity of the gel: $f_{cross} = G/\eta_s$. The lower limit in eqn (6) draws a condition on the capillary number of the substrate,

$$Ca_s = \frac{\eta_s v}{\gamma_s} \approx 1. \quad (7)$$

It implies that the continuous/stick–slip transition at a low $f$ is determined by which force dominates the system, i.e., viscous forces in the substrate or capillary forces.

In order to check our theoretical prediction, we superposed the expected critical values for the boundaries of stick–slip regime $f_{cross} \, l_e/l_{cap}$ and $f_{cross}$ in Fig. 9. As expected, the domain of the regular stick–slip is entirely contained between the two limits.

The lower limit is higher than expected, which suggests that some prefactors should be considered in our analysis. One of the possibilities for this prefactor is the asymmetric profile of the surface ridge, as it was discussed by Bostwick et al.[42] Since the receding contact angle on the SBS–paraffin gel is smaller than 90° (Fig. 4), the asymmetric profile of the elasto-capillary deformation and the unbalance between the interfacial tensions $\gamma_s$ and $\gamma_{sl}$ could suggest a closer bound.

Finally, we mention that our qualitative interpretation is reminiscent of a theoretical analysis of wetting on a visco-elastic substrate, recently reported by Karpitschka et al.[43] The same kind of continuous band of spatial mode is involved in the surface deformation, and a condition on the substrate capillary number is also found at the end. It would be interesting to compare both approaches for further theoretical development.

## V. Conclusions

In this article, we have investigated the dynamics of a liquid contact line moving on a hydrophobic and visco-elastic (SBS–paraffin) gel substrate in a dip-coating geometry. Depending on the translation velocity of the substrate $v$, the contact line exhibits different regimes of motions, i.e., continuous motion at a high $v$, stick–slip motion at an intermediate $v$ and another continuous motion at a low $v$. Furthermore, the stick–slip motion can be classified into two different kinds: large and periodic jumps occurring coherently in the entire part of the contact line and irregular stick–slip by localized pinning spots.

We have conjectured that such transitions are determined by the competition between different frequencies, i.e., the frequency $f$ of the surface deformation induced by the moving contact line and the gel crossover frequency $f_{cross} = 1/\tau_{gel}$ related to the material relaxation. In particular, we have found that all the threshold values of $f$ between different regimes are proportional to $f_{cross}$.

Finally, we have proposed a qualitative modelling which predicts the existing range of the stick–slip regimes. Therein, we considered the continuous spectrum associated with the surface deformation that ranges from the meniscus size $l_{cap}$ to the elasto-capillary length $l_e$: $1/l_{cap} < k < 1/l_e$. Our model grasps essential features of the observed phenomena.

## Acknowledgements

The funding of this project has been covered by a grant from labex SEAM (Science and Engineering for Advanced Materials) with reference ANR-11-LABX-086 of the program "Future Investment" with reference ANR-11-IDEX-0005-02 administrated by the French National Agency for Research (ANR).